\documentclass[preprint,preprintnumbers,amsmath,amssymb]{revtex4}
\usepackage{graphicx}
\usepackage{dcolumn}
\usepackage{bm}
\begin{document}
\title{Massive spinor fields in flat space-times with non-trivial topology}
\author{N.~Ahmadi$^{a}\thanks{Electronic address:~nahmadi@ut.ac.ir}$ and
M.~Nouri-Zonoz$^{a,b}\thanks{Electronic address:~nouri@theory.ipm.ac.ir}$ }
\address{$^{a} $ Department of Physics, University of Tehran, End of North Karegar St.,
Tehran 14395-547, Iran.\\
$^{b}$Institute for studies in theoretical physics and
mathematics, P O Box 19395-5531 Tehran, Iran.}
\begin{abstract}
The vacuum expectation value of the stress-energy tensor is calculated for spin $1\over 2$ massive fields in several multiply connected flat space-times. We examine the physical effects of topology on manifolds such as $R^3 \times S^1$, $R^2\times T^2$, $R^1 \times T^3$, the Mobius strip and the Klein bottle. We find that the spinor vacuum stress tensor has the opposite sign to, and twice the magnitude of, the scalar tensor in orientable manifolds. Extending the above considerations to the case of Misner space-time, we calculate the vacuum expectation value of spinor stress-energy tensor in this space and discuss its implications for the chronology protection conjecture.
\end{abstract}
\maketitle
\section{Introduction}
The topology of a space-time manifold $\cal M$ can have important implications for the construction of classical and quantum fields. To see the physical effects of topology, one could relax the curvature effects by studying quantized fields in multiply connected {\it flat} space-times. Field theory in space-times carrying a non-trivial fundamental group (authomorphic fields) has been studied previously through different approaches [1-4]. The
effect of of mass is studied for scalar fields in the context of Casimir energy in toroidal space-times [5]. The effects of multiple connectedness of the space-time manifolds for massless fields in various topological spaces are studied by Dewitt, Hart and Isham [4]. Following their work, Tanaka and Hiscock [6] evaluated vacuum expectation value of the stress tensor $<0|T_{\mu\nu}|0>$  for free massive scalar fields in four dimensional space-time manifolds of the type $R^1(time)\times \Sigma (space^3)$. \\
In what follows we extend the above studies to the case of free massive spinor fields. The Minkowski vacuum state is assumed to be the natural vacuum in all space-times considered. We shall describe the expectation value of the stress tensor $T_{\mu\nu}$ in this vacuum state and in case there is no global timelike Killing vector field, as in Misner space-time, we will discuss how the default vacuum state could be taken to be that of the Minkowski space-time [7].\\
Misner space has closed timelike curves (CTCs) in bounded regions (nonchronal regions) of the space-time which are separated from the CTC free regions (chronal regions) by a special type of Cauchy horizon the so called {\it chronology horizon}, the null surface beyond which CTCs first form. It is believed that nature will not allow the formation of CTCs and this is embodied in Hawking's "chronology protection conjecture:The laws of Physics do not allow the appearance of closed timelike curves" [8]. The main impediment found to the appearance of such curves is the divergence of the vacuum stress-energy tensor of the quantized fields on the chronology horizon. It is believed that the gravitational backreaction to these diverging stress-energy tensor would alter the space-time in such a way as to prevent the formation of CTCs. In previous studies [6-13], it has been shown that the divergence of $<0|T_{\mu\nu}|0>$ for massless scalar fields on various space-times with chronology horizons cannot be generally avoided. Here we will show that the same thing is true for massive spinor field on Misner space. \\
The outline of the paper is as follows: In section II we review the spin structure in curved space-times. A general procedure for calculating the vacuum expectation value of $T_{\mu\nu}$ for a free massive spinor field in a flat multiply connected space-time is described in section III. We apply this method to orientable and non-orientable space-times with different spin structures in section IV. The vacuum stress-energy tensor for  massive spinor fields and their massless limit in Misner space are discussed in section V. Conclusions will be discussed in the last section.
\section{Spin structures}
To define a spinor field on a manifold $\cal M$, it must be equipped with a Riemannian (or pseudo Riemannian) metric, together with a bundle of orthonormal frames $\left\{{\bf e}_U ,{\bf e}_V,....\right\}$  over the patches $\left\{ U,V,...\right\}$ which are employed giving the bundle a Lorentzian structure group. In an overlap we shall assume ${\bf e}_V(x)= {\bf e}_U(x)C_{UV}(x)$ , where $C_{UV}(x)$, $x \in U \bigcap V$ is the transition function, $C_{UV}(x):U \bigcap V \rightarrow L_0 $ in which $L_0$ is the Lorentz transformation which preserves the direction of time. The transition functions $C_{UV}$ certainly satisfy the consistency condition
$$C_{UV}C_{VW}C_{WU}=1\;\;\;\;\; , \;\;\;\;\; C_{UU} = 1\eqno(1)$$
A spin structure on a Lorentzian 4-manifold ${\cal M}$ is defined by the transition function $C^\prime_{UV}  \in SL(2,C)$ such that $\Lambda (C^\prime_{UV}) = C_{UV}(x)$, where $\Lambda  : SL(2,C) \rightarrow L_0$ is the double covering of the Lorentz group. Because of the ambiguity in choosing any of the two coverings $\pm A \in SL(2,C)$, it is not clear that $C^\prime_{UV}$ can be chosen consistently to satisfy the consistency conditions (1) as $C_{UV}$ does.
If this can be done, the structure group of the tangent bundle of ${\cal M}$ is said to be {\it lifted}  from the Lorentz group to the $SL(2,C)$ group and ${\cal M}$ is equipped with a spin structure. Choosing different frames ${\bf e}_U$  in $U$ could in principle lead to different spin structures. To see this, suppose that $p$ is in a patch $U$ covered by a Lorentzian frame field ${\bf e}_U$. If we take the frame ${\bf f}(p)={\bf e}_U(p)$ and transport it around any closed curve $C(t)$, $0<t<1$  lying in $U$, on returning to the same frame ${\bf f}(p)$ we can compare ${\bf f}(C(t))$ with ${\bf f}(p)= {\bf f}(C(0))$ as follows: Identify all frames ${\bf e}_U$ at points of $U$ with the single frame ${\bf e}_U$ at $p$. Then by comparing  ${\bf f}(C(t))$ with ${\bf e}_U(C(t))$ we find ${\bf f}(C(t)) = {\bf e}_U(C(t))\Lambda(t)$, in other words we have traced out a closed curve $t\rightarrow \Lambda(t)$ in Lorentz group. If ${\cal M}$ has a spin structure, i.e $SL(2,C)$ is its tangent bundle structure group, and we transport a frame $\bf f$ around any closed path $C$ in ${\cal M}$, upon returning to the same Lorentzian frame we can decide whether the frame has made an even or an odd number of complete rotations!. This is so, because by considering the $SL(2,C)$ frame bundle to ${\cal M}$, the curve $C$ in ${\cal M}$ is covered by a unique curve in this frame bundle starting at $I$, defined by $\bf f$. Upon returning to the starting point of $C$, the lifted curve will return to its starting point, corresponding to an even number of rotations, or to a point in the frame bundle related to the initial point by $-I\in SL(2,C)$, corresponding to an odd number of rotations [14].\\
Not all manifolds admit spin structure. In fact the necessary and sufficient condition for the existence of such a structure is the topological restriction that the second Stiffel-Whitney class of $\cal M$, $ H_2 ({\cal M},Z_2)$, vanishes [4,14].
If $\cal M$ has a spin structute, the Dirac spinor bundle is simply the vector bundle associated with the $SL(2,C)$ tangent bundle through the $4\times 4$ representation $\rho$ of $SL(2,C)$,
$$\rho_{UV}(x) = \rho(C^\prime _{UV}(x))=
\left[ \begin{matrix}
 C^\prime_{UV}(x) & 0 \cr 0 & C^{\prime
{-1}\dagger}_{UV}(x) \cr
\end{matrix}
\right]\eqno(2)$$ 
This spinor bundle is the bundle whose cross sections $\Psi$ are known as spinor
fields (wave functions). As $SL(2,C)$ spin frame bundle is
trivial, existence of different spin structures does not lead
automatically to twisted spinor fields. Instead this fact reflects itself in the
different spin connections {\it pulled back} by the different maps
$\Lambda$ to give a connection $\Lambda_\ast\omega$ on $SL(2,c)$
tangent bundle, where $\omega$ is the connection form for the
Lorentzian tangent bundle. The associated connection in spinor
bundle is given by $\Omega = \rho_{\ast}(\Lambda_\ast\omega)$,
which is employed to construct covariant derivative of spinor
fields in the spinor bundle. It is also shown that the Lagrangian
associated with different spin structures cannot be made equal by
a spinor field gauge transformation and lead to different physical
results [4].
\section{Calculation of $<0|T_{\mu\nu}|0>$ in a multiply connected flat space-time}

In a curved space, the action and stress-energy for a free spinor field $\Psi$ are given by
\footnote{For simplicity and to enable comparison with previous results, particularly those 
obtained in [4], we assume the spinor field to be neutral and represented by a Majorana spinor.};
 $$S \left[ \Psi \right] = {i\over 2}\int g^{1/2}(\bar{\Psi}\gamma^\rho \Psi_{;\rho} - \bar{\Psi}_{;\rho}\gamma^\rho\Psi -M\bar{\Psi}\Psi) d^4 x \eqno(3)$$
and
$$ T^{\mu\nu} = -{1\over 4} i \left[(\bar{\Psi}\gamma^\mu \Psi^{;\nu} +  \bar{\Psi}\gamma^\nu \Psi^{;\mu}) - (\bar{\Psi}^{;\mu}\gamma^\nu \Psi + \bar{\Psi}^{;\nu}\gamma^\mu \Psi)\right]\eqno(4)$$
the latter of which could be written in the following form,
$$ T^{\mu\nu} = {1\over 4}i {\rm Tr}(\gamma^{( \mu}\left[\Psi^{;\nu )},\bar{\Psi}\right] - \gamma^{(\mu}\left[{\Psi},\bar {\Psi}^{;\nu )}\right])\eqno(5)$$
where $A_{(\mu}B_{\nu)}={1\over 2}(A_\mu B_\nu + A_\nu B_\mu )$ and trace is over the suppressed spinor indices. The transition from classical to quantum fields is made by replacing the classical fields by field operators. These quantities diverge when their vacuum expectation values are taken. To obtain the finite, physical contribution to $<0|T_{\mu\nu}|0>$, we employ the manifestly covariant {\it point separation} regularization method [15-16]. Based on Schwinger's proper time technique, this method gives equivalent results to other methods such as the dimensional and zeta-function regularization schemes. Each term in (5) is constructed out of products of either field operators or their derivatives at {\it the same space-time point}. Taking this to be the prime source of divergent terms, the calculations in this method are done by first moving one operator in each product to a nearby point and then let them to coincide at the end. In other words, we split the point $x$ into $x$ and $\tilde{x}$ and take the coincidence limit $\tilde{x} \rightarrow x$, e.g. for a typical commutation relation in (5) we have,
$$ \left[\Psi^{;\nu},\bar{\Psi}\right] = \lim_{\tilde{x}\rightarrow x}{1\over 2} \left\{ \left[\Psi^{;\nu^\prime},\bar{\Psi}\right] + \left[\Psi^{;\nu},\bar{\Psi}\right] \right\}\eqno(6)$$
The primed derivatives are taken with respect to $\tilde{x}$. Expressing the point-separated $<0|T_{\mu\nu}|0>$ in terms of the so called spinor Hadamard elementary functions,
$$S^{(1)}_{\alpha\beta} (x,\tilde{x}) = <0|\left[ \Psi_\alpha(x) , \bar{\Psi}_\beta (\tilde{x})\right]|0> \eqno(7)$$
we find
$$<0|T_{\mu\nu}|0> = {1\over 8}i \lim_{\tilde{x}\rightarrow x} {\rm Tr}\gamma^{(\mu}\left(S^{(1);\nu)} -S^{(1);\nu^{\prime})}\right)\eqno(8)$$
The spinor Hadamard function, on the other hand, could be written in terms of the scalar Hadamard function as
$$S^{(1)}(x,\tilde{x}) = -\left( i\gamma^\rho G_{;\rho}^{(1)} + MG^{(1)}\right) (x,\tilde{x})\eqno(9)$$
in which $G^{(1)}$ satisfies $(\Box - M^2)G^{(1)}(x,\tilde{x}) = 0$ .
Therefore, (8) takes the following new form,
$$<0|T_{\mu\nu}|0> = {1\over 8}\lim_{\tilde{x}\rightarrow x} {\rm Tr}\gamma^{(\mu}\gamma^\rho \left( G_{;\rho}^{(1)\nu)}-G_{;\rho}^{(1)\nu^\prime)}\right)(x,\tilde{x})\;\;\;\;\;\;\;\;\;\;\;\;\;\;\;\;\;\;\;\;\;\;\;\;\;\;\;\;\;\;\;\;\;\;\;\;\;\;\;\;\;$$
$$\;\;\;\;\;\;\;\;\;\;\;\;\;\;\;\;= {1\over 16}\lim_{\tilde{x}\rightarrow x} {\rm Tr}\left[(\gamma^{\mu}\nabla^\nu + \gamma^{\nu}\nabla^\mu) - (\gamma^{\mu}\tilde{\nabla}^\nu + \gamma^{\nu}\tilde{\nabla}^\mu)\right] \gamma^\sigma \nabla_\sigma G^{(1)}(x,\tilde{x}) \eqno(10)$$
The Hadamard function for a massive scalar field in Minkowski space could be written as a function of the half squared geodesic distance, $\sigma = {1\over 2}g_{\alpha\beta}(x^\alpha - \tilde{x}^\alpha)( x^\beta - \tilde{x}^\beta)$, between two points $x$  and $\tilde{x}$ in the form of [17],
$$G_0^{(1)}(x,\tilde{x}) = {M \over {2\pi^2 \sqrt{2\sigma}}}\Theta(2\sigma)K_1(M\sqrt{2\sigma}) + {M \over {4\pi \sqrt{-2\sigma}}}\Theta(-2\sigma)I_1(M\sqrt{-2\sigma})\eqno(11)$$
where $\Theta$  is a step function and $k_1$ and $I_1$ are the modified Bessel functions of the first and second kind respectively.
Since the space-times we are going to consider are flat multiply connected, each could be constructed by a topological identification out of Minkowski space with metric given by
$${\rm ds}^2 = -({\rm d}x_0)^2 + ({\rm d}x_1)^2 + ({\rm d}x_2)^2 + ({\rm d}x_3)^2 \eqno(12)$$
Consequently field theory in them could be built up from field theory in Minkowski space-time by using the method of images. In this method, all inequivalent spacelike paths connecting two point $x$ and $\tilde{x}$ are taken into account, for they cannot be deformed continuously to each other. The images of $\tilde{x}$, which we label them with integer $n$ and are located at $\tilde{x}+ na$, are connected to the point $x$ through $\sigma = {1\over 2}g_{\alpha\beta}(x^\alpha - \tilde{x}_n^\alpha)( x^\beta - \tilde{x}_n^\beta)$. The renormalized vacuum stress-energy tensor, $<0|T^{\mu\nu}|0>$, is given by the renormalized Hadamard function $G_{ren}^{(1)}$, which has a contribution from each image charge, except that of $\tilde{x_0}$ which is divergent and is associated with the Minkowski vacuum state, i.e.,
$$G_{ren}^{(1)}(x,\tilde{x}) = \sum_{{n=-\infty}\atop{n\neq 0}}^\infty G_0^{(1)}(x,\tilde{x}_n)\eqno(13)$$
One can obtain an algebraic expression for $<0|T^{\mu\nu}|0>$ by following the steps below,\\
- Finding an appropriate identification of the points in Minkowski space,\\
-   Choosing local frames on $\cal M$ ,\\
-   Since the frames in $x$ and $x_n$ are related through a Lorentzian transformation, the frame in  may not be the usual Minkowski frame. As spinor fields obey the appropriate transformation law under change of frame, it translates into the spinors as $\Psi(x_n)= S \Psi(x)$, where $\Psi(x_n)$ is the spinor at $x_n$ with respect to the transformed frame. Topological identifications demand us to impose a suitable condition on spinor fields, i.e. $\Psi(x)= S^{-1} \Psi(x_n)$, which simply compensates for the transformation, and  makes covariant and ordinary derivatives coincide,\\
-   Computing the derivatives, \\
-   And taking the limit as the separated points are brought together.
\section{Massive spinor fields in orientable manifolds}
By following the above steps, the vacuum expectation value $<0|T^{\mu\nu}|0>$ of the stress-energy tensor of a free massive spinor field is evaluated in four dimensional orientable space-times with $R(time)\times \Sigma(space^3)$ topology. We try to find every possible spin structure in each case, including the twisted and untwisted spin connections.
\subsection{Untwisted spin connnection}
The first topology to consider is $\Sigma = S^1 \times R^2$. The periodicity represented by $S^1$ is oriented in spatial $x^1$ direction, and the identical points in this topology are
$$(x^0, x^1, x^2, x^3 )\leftrightarrow (x^0, x^1+na, x^2, x^3 )\eqno(14)$$
where $n$ is an integer and Cartesian coordinates $(x^0, x^1, x^2, x^3)$  are used in ${\cal M}$  . The half squared geodesic distance, $\sigma_n$, between the point $x$ and the $n$-th image charge at $\tilde{x}$ is given by
$$ \sigma_n = {1\over 2}\left[-(x^0 - \tilde{x}^0)^2 + (x^1 - \tilde{x}^1-na)^2 + (x^2 - \tilde{x}^2)^2 +(x^3 - \tilde{x}^3)^2\right]\eqno(15)$$
Due to the fact that in all space-times considered we have periodicity in spatial section of $\cal M$, the intervals between the image charges are always spacelike, that is $\sigma_n>0$ and consequently only the first term in (11) will concern us.
We choose the local frames to be everywhere parallel to the Cartesian axis. We will choose such a local frame to obtain untwisted spin connections in all space-times considered.
One can see that in orientable space-times in which the geodesic distance between the points $x$ and $\tilde{x}_n$ is a function of $x^\alpha - \tilde{x}^\alpha_n$, we have $\tilde{\nabla}_\mu G_{ren}^{(1)}(x,\tilde{x})= -\nabla_\mu G_{ren}^{(1)}(x,\tilde{x})$, and equation (10) reduces to
$$<0|T^{\mu\nu}|0> = \lim_{\tilde{x}\rightarrow x}\nabla^\mu \nabla^\nu G_{ren}^{(1)}(x,\tilde{x})\eqno(16)$$
So that the vacuum stress tensor for untwisted massive spinor fields on $S^1 \times R^3$ is
$$<0|T^{\mu\nu}|0> = -{M^4\over \pi^2} \sum_{n=1}^\infty \left\{{K_2(z_{n})\over z_{n}^2}g_{\mu\nu} + {K_3(z_{n})\over z_{n}}diag[0, -1, 0, 0]\right\} \eqno(17)$$
where $z_n = |Mna|$. The vacuum stress tensor is seen to have the opposite sign, and twice the magnitude of, the scalar stress-energy tensor given in reference [6] (equation (11)). The sign and magnitude changes are respectively due to fermionic statistics and degrees of freedom. Exactly the same relation holds between vacuum expectation values of massless scalar and spinor fields [4]. 
The next natural step would be to consider space-times with spatial topology $\Sigma = T^2 \times R^1$,  which are closed in two directions $x^1$ and $x^2$. The identified points are (again in the Cartesian coordinates)
$$(x^0, x^1, x^2, x^3 )\leftrightarrow (x^0, x^1+na, x^2+mb, x^3 )\eqno(18)$$
The $n$-th image charge of $\tilde{x}$ is located at $\tilde{x}_{nm}$  and the half squared geodesic distance between point $x$ and $\tilde{x}_{nm}$ is given by
$$ \sigma_n = {1\over 2}\left[-(x^0 - \tilde{x}^0)^2 + (x^1 - \tilde{x}^1-na)^2 + (x^2 - \tilde{x}^2-mb)^2 +(x^3 - \tilde{x}^3)^2\right]\eqno(19)$$
Again we can write $\tilde{\nabla}_\mu G_{ren}^{(1)}(x,\tilde{x})= -\nabla_\mu G_{ren}^{(1)}(x,\tilde{x})$, and for untwisted spin connection we have
$$<0|T^{\mu\nu}|0> = -{M^4\over 2\pi^2} \sum_{{n,m=-\infty}\atop{(m,n)\neq (0,0)}}^\infty \left\{{K_2(z_{nm})\over z_{nm}^2}g_{\mu\nu} + M^2{K_3(z_{nm})\over z_{nm}^3}diag[0, -n^2a^2, -m^2b^2, 0]\right\} \eqno(20)$$
where $z_{nm} = M(n^2a^2 + m^2b^2)^{1/2}$.
The last example of the space-times whose stress-energy tensor for free spinor fields can be read off that of scalar fields is a space-time with spatial topology $\Sigma = T^3$ obtained through the following identification of Minkowski space
$$(x^0, x^1, x^2, x^3 )\leftrightarrow (x^0, x^1+na, x^2+mb, x^3+lc )\eqno(21)$$
where $n,m$ and $l$ are integers and $a,b$ and $c$ are the periodicities in the $x^1, x^2$ and $x^3$ directions, respectively. The half squared geodesic distance, $\sigma_{nml}$, is given by
$$ \sigma_{nml} = {1\over 2}\left[-(x^0 - \tilde{x}^0)^2 + (x^1 - \tilde{x}^1-na)^2 + (x^2 - \tilde{x}^2-mb)^2 +(x^3 - \tilde{x}^3-lc)^2\right]\eqno(22)$$
and the resulting vacuum expectation value of stress-energy tensor is
$$<0|T^{\mu\nu}|0> = -{M^4\over 2\pi^2} \sum_{{n,m,l=-\infty}\atop{(m,n,l)\neq (0,0,0)}}^\infty \left\{{K_2(z_{nml})\over z_{nml}^2}g_{\mu\nu} + M^2{K_3(z_{nm})\over z_{nml}^3}diag[0, -n^2a^2, -m^2b^2, -l^2c^2]\right\} \eqno(23)$$
where $z_{nml}= M(n^2a^2 + m^2b^2 + l^2c^2)^{1/2}$.
Comparing equations (20) and (23) with those obtained for massive scalar fields in [6] we note that the same relation holds between their signs and magnitudes as in the case of  $S^1 \times R^3$ topology.
\begin{figure}
\begin{center}
\includegraphics[angle=-90,scale=0.7]{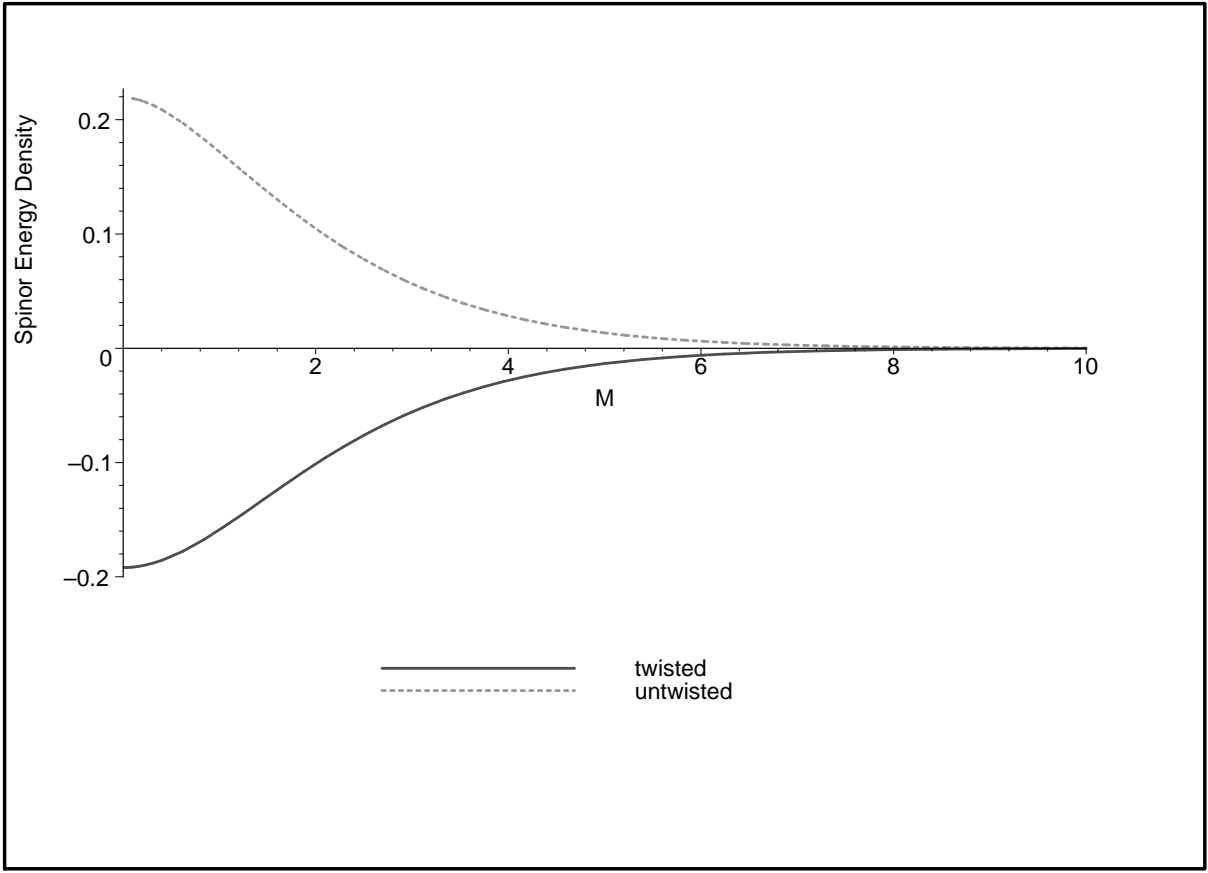}
\caption{Plots of the energy densities vs. the field mass $M$ with periodicity $a=1$ for the $\Sigma =s^1 \times R^2$ topology.}
\end{center}
\end{figure}
\newpage
\subsection{Twisted spin connections}
Twisted spin connections in the preceding topologies of are  simply obtained by choosing local frames on $\cal M$ such that they rotate smoothly through $2\pi$ about some chosen axis as one goes from $x$ to its image. To {\it undo} the effect of rotation {\it locally}, the anti-periodicity conditions are forced in each case such that the odd and even modes appear with different signs. In the simplest case of $\Sigma = S^1\times R^2$  the antiperiodicity on $\Psi(x)$ is given by,
$$\Psi(x^0, x^1, x^2, x^3) = (-1)^n \Psi(x^0, x^1+na, x^2, x^3)\eqno(24)$$
The same antiperiodicity condition is introduced into $G^{(1)}_{ren}(x,\tilde{x})$ by inserting the factor of $(-1)^n$ in (13) such that its twisted version is given by
$$G_{{ren}\atop{twisted}}^{(1)}(\sigma) = \sum_{{n=-\infty}\atop{n \neq 0}}^\infty (-1)^n G_0^{(1)}(\sigma_n)\eqno(25)$$
Three inequivalent twisted connections are introduced on $T^2\times R^1$ by inserting $(-1)^n, (-1)^{m}$ and $(-1)^{n+m}$ into the related summand. Similarly, we have seven different twisted connections in  $T^3$, whose corresponding expectation values are obtained by inserting $(-1)^n, (-1)^{m}$ and $(-1)^{l}$ or any combination of them into the summand.
All the space-times considered are orientable and it is the connection that can be twisted, not the spinor fields themselves.
\begin{figure}
\begin{center}
\includegraphics[angle=-90,scale=0.7]{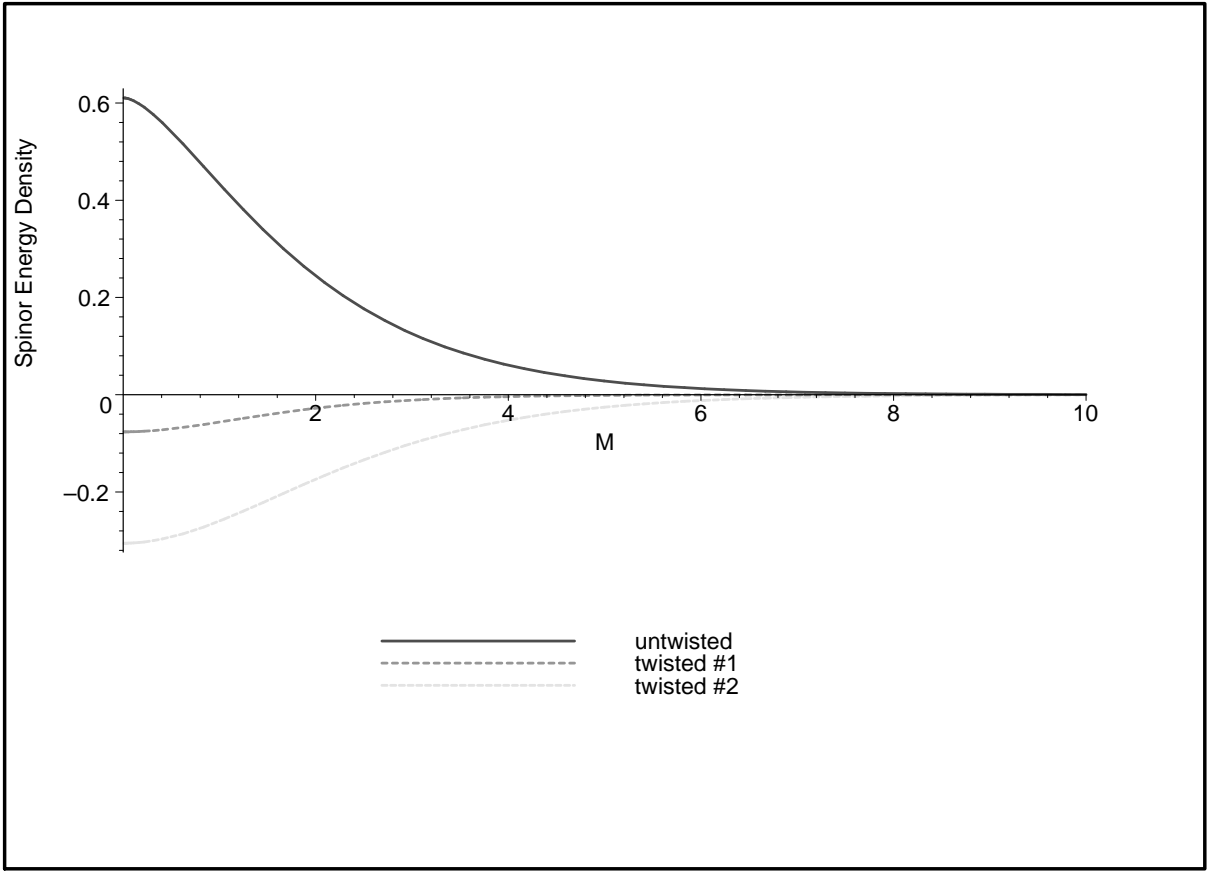}
\caption{Plots of the energy densities vs. the field mass $M$ with periodicities $a=b=1$ for the $\Sigma =T^2 \times R^1$ topology.}
\end{center}
\end{figure}
\begin{figure}
\begin{center}
\includegraphics[angle=-90,scale=0.7]{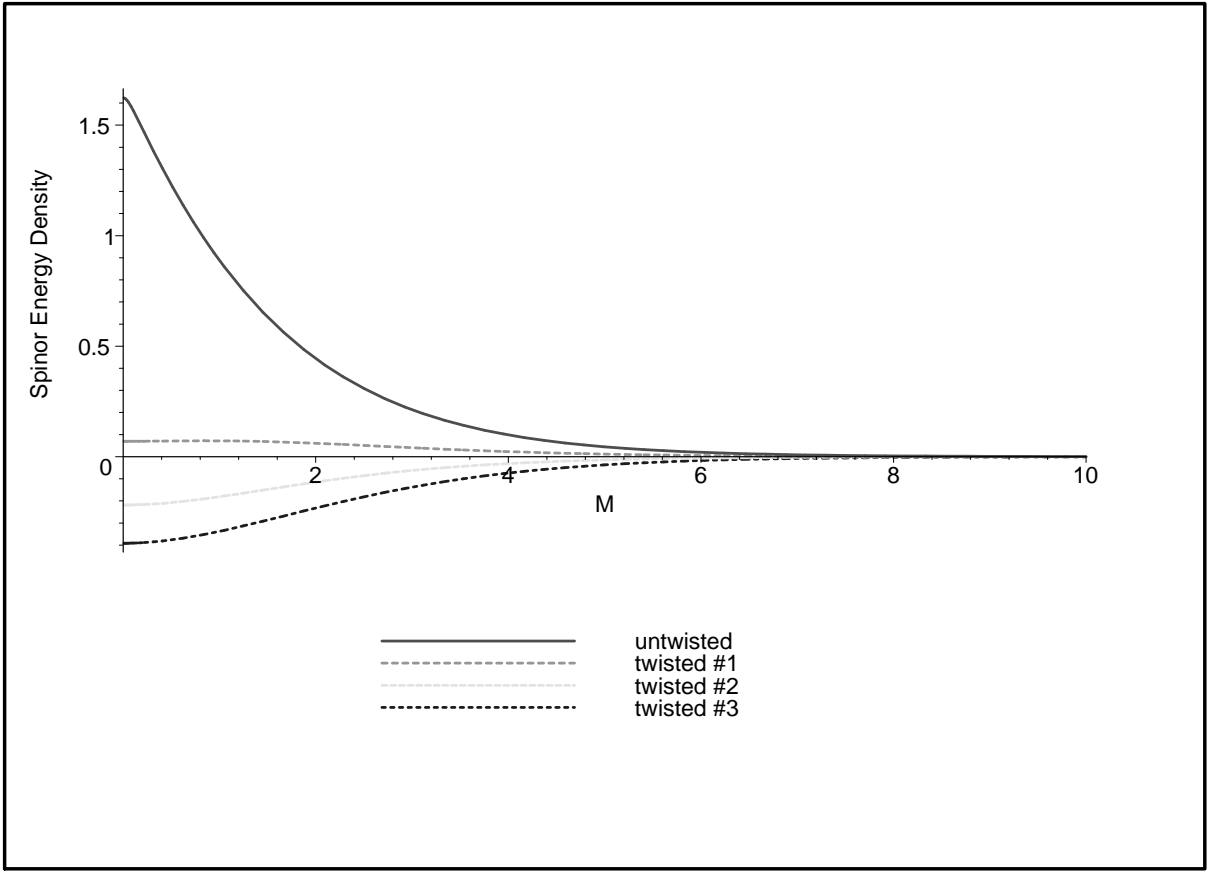}
\caption{Plots of the energy densities vs. the field mass $M$ with periodicities $a=b=c=1$ for the $\Sigma =T^3 $ topology.}
\end{center}
\end{figure}
As discussed in [18], for static space-times $<0|T^{0i}|0>=0$. The usual covariance arguments require that $<0|T^{\mu\nu}|0>$ be proportional to the Minkowski metric, however, point identification in Minkowski space-time will modify the topological properties of infinite flat space-time, which in general, will destroy global Poincare invariance. Of course, it will exist as a local invariance but this is not enough.
Plots of energy density $\rho = <0|T^{00}|0>$ as a function of the field mass $M$ for the untwisted and twisted connections in $S^1 \times R^3$, $T^2\times R^2$ and $T^3\times R^1$ topologies are shown in figures 1-3 respectively. Compared to the massless case, the effect of mass on the energy density is to reduce its magnitude. The energy density magnitudes in the massless $S^1\times R^3$ case are obtained from (17) in the limit $M \rightarrow 0$ as,
$$\sum_{n=1}^\infty {2\over {\pi^2(z_n/M)^4}} = {\pi^2 \over 45{a^4}}\eqno(26)$$
$$\sum_{n=1}^\infty {2(-1)^n\over {\pi^2(z_n/M)^4}} = -{7\over 8}{\pi^2 \over 45{a^4}}\eqno(27)$$
for the untwisted and twisted spin connections respectively. These are in complete agreement with the results obtained in [4]. An interesting feature in the twisted $T^3$ case is that the energy density could be positive and for equal periodicities its maximum is shifted from $M=0$ (Fig. 3).
\section{Massive spinor fields in Non-orientable manifolds}
The non-orientability of manifold affects the spin structure, in that the local frames cannot be defined globally. In a non-orientable manifold, transport around the closed spatial directions or going from  $x$ to any of its images may change the handedness of the local frame. Therefore, imposing a condition on $\Psi(x)$ is necessary to construct a consistent spinor structure on these manifolds.
A four dimensional Mobius strip $R^1 \times \Sigma = R^1 \times M^2 \times R^1$ is a non-orientable space-time manifold which can be constructed by the following identification in Minkowski space-time
$$(x^0, x^1, x^2, x^3 )\leftrightarrow (x^0, x^1+na,(-1)^n x^2, x^3 )\eqno(28)$$
The half squared geodesic distance $\sigma_n$ is equal to
$$ \sigma_n = {1\over 2}\left[-(x^0 - \tilde{x}^0)^2 + (x^1 - \tilde{x}^1-na)^2 + (x^2 - (-1)^n\tilde{x}^2)^2 +(x^3 - \tilde{x}^3)^2\right]\eqno(29)$$
If we choose local frames parallel to the coordinate axes, with going from $x$ to $x_n$ ($n$ odd), the direction of the local $x^2$ axis reverses. Since the transformation that relates the local frame in $x$ to the one in $x_n$ is parity, a suitable condition on spinor fields must relate $\Psi(x)$ to $\Psi(x_n)$ through a transformation $S$ that induces the reversal of $x^2$ as the magnitude of the coordinate $x^1$ increases by $na$. Therefore $S$ should be a $4\times 4$ representation of $SL(2,C)$ such that $S^2=I$.
By the above discussion the condition
$$\Psi(x^0, x^1, x^2, x^3 ) = (\gamma^0\gamma^3\gamma^1)^n \Psi(x^0, x^1+na, (-1)^n x^2, x^3 )\eqno(30)$$
must be considered for all integers $n$, and we have $(\gamma^0\gamma^3\gamma^1)^2=I$. The spinor stress-energy tensor may now be computed from equation (10), where, in view of the above condition, we have
$$G_{ren}^{(1)}(\sigma) = \sum_{{n=-\infty}\atop{n \neq 0}}^\infty \left[ G_0^{(1)}(\sigma_{2n}) + G_0^{(1)}(\sigma_{2n+1})(\gamma^0\gamma^3\gamma^1)\right]\eqno(31)$$
Since the trace of an odd number of  $\gamma$- matrices vanishes, taking the trace reveals that the second sum in equation (31) make no contribution and we immediately regain the result given for the topology , $\Sigma=S^1 \times R^2$ with periodicity $2a$ in the $x^1$ direction, i.e., we will have the same result as in (17) but now with $z_n=|2Mna|$. This is not unexpected as the Mobius strip manifold is locally $S^1 \times R^3$. It is seen that twisted spin connection has no effect on the stress-energy tensor, as it just changes the sign of the second term in (31).
The next example of a non-orientable manifold is the Klein bottle, $\Sigma=K^2 \times R^1$, which is obtained by the following identification of points in Minkowski space,
$$(x^0, x^1, x^2, x^3 )\leftrightarrow (x^0, x^1+na,(-1)^n x^2+2mb, x^3 )\eqno(32)$$
where now the half squared geodesic distance is given by,
$$ \sigma_{nm} = {1\over 2}\left[-(x^0 - \tilde{x}^0)^2 + (x^1 - \tilde{x}^1-na)^2 + (x^2 - (-1)^n\tilde{x}^2-2mb)^2 +(x^3 - \tilde{x}^3)^2\right]\eqno(33)$$
Again, due to the non-orientability of Klein bottle, we impose the following condition on $\Psi(x)$,
$$\Psi(x^0, x^1, x^2, x^3 ) = (\gamma^0\gamma^3\gamma^1)^n \Psi(x^0, x^1+na, (-1)^n x^2+2mb, x^3 )\eqno(34)$$
As in the case of Mobius strip, by taking the trace, the odd $n$ terms make no contribution. Therefore, neither the Mobiosity nor twisted spin connection in the $x^1$ direction, will affect  spinor vacuum stress tensor and we obtain,
$$<0|T^{\mu\nu}|0> = -{M^4\over 2\pi^2} \sum_{{n,m=-\infty}\atop{(m,n)\neq (0,0)}}^\infty \left\{{K_2(z_{nm})\over z_{nm}^2}g_{\mu\nu} + M^2{K_3(z_{nm})\over z^3_{nm}}diag[0, -4n^2a^2, -4m^2b^2, 0]\right\} \eqno(35)$$
where $z_{nm}= M[n^2(2a)^2 + m^2(2b)^2]^{1/2}$, which is identical to that for $R^2 \times T^2$, with spatial periodicities doubled in both directions. In the massless limit it may now be compared with that  obtained in [4]. Although $\sigma_{nm}$ depends on $\tilde{x}^2$, even after the coincidence limit is taken, (35) is a coordinate independent result. This is in sharp contrast to the results obtained in the massive scalar case where energy density contains coordinate dependent terms, oscillatory in the $x^2$ direction [6]. This feature is retained in the twisted case as the twisted spin connection in the $x^2$ direction only inserts the extra factor $(-1)^m$ in the summand. Plots of the untwisted energy densities for orintable and nonorientable manifolds are shown in Fig. 4.
\begin{figure}
\begin{center}
\includegraphics[angle=-90,scale=0.7]{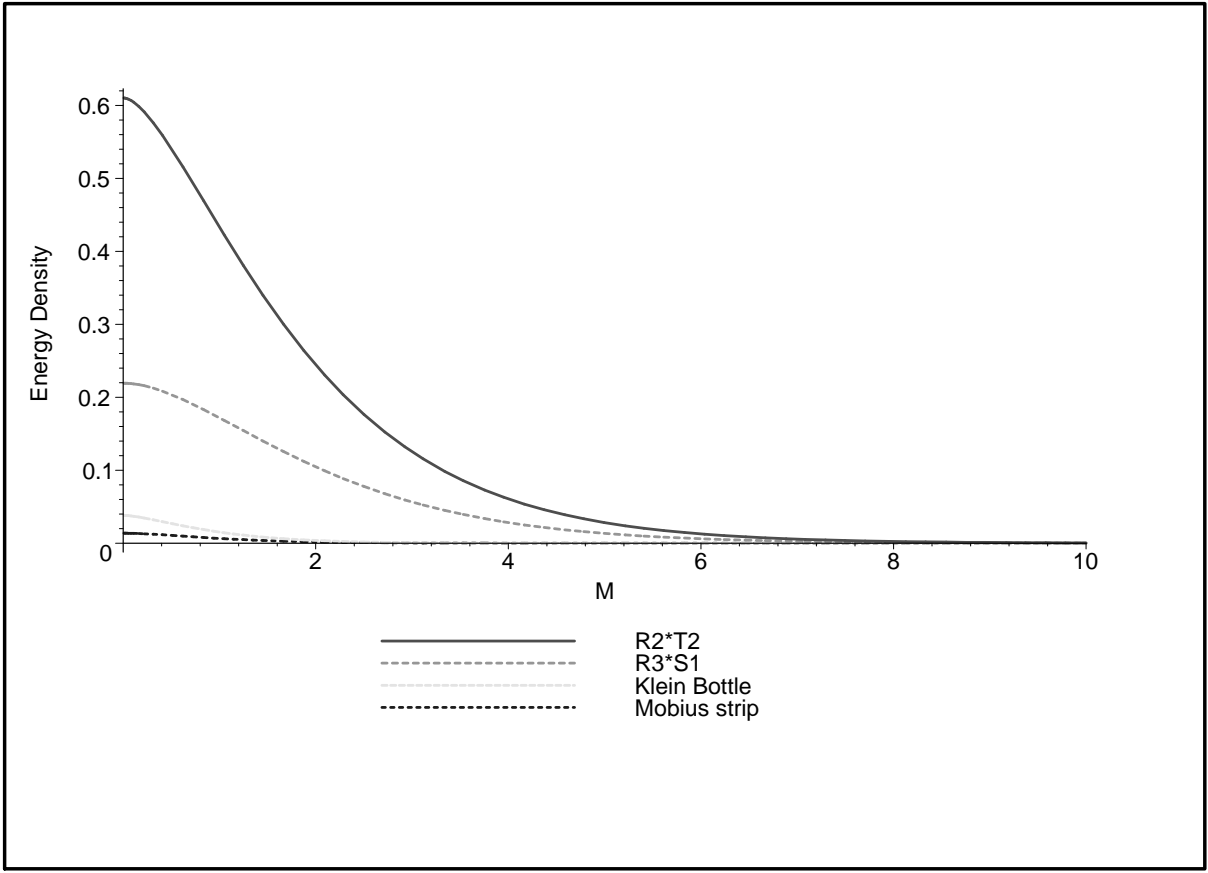}
\caption{Plot of the untwisted energy densities  vs. the field mass $M$ with periodicities $a=b=c=1$ for orientable and nonorientable topologies.}
\end{center}
\end{figure}
\section{Spinor fields in Misner space}
Misner space is somewhat different from the space-times we have examined so far, though it shares some of their characteristics which makes it an interesting case for the calculation of $<0|T^{\mu\nu}|0>$. It has flat Kasner metric and $S^1 \times R^3$ topology. In the Misner coordinates, $(y^0, y^1, y^2, y^3)$, the metric is given by,
$${\rm ds}^2 = -({\rm d}y^0)^2 + ({y^0})^2({\rm d}y^1)^2 + ({\rm d}y^2)^2 + ({\rm d}y^3)^2 \eqno(36)$$
The flatness of Misner space becomes obvious by transforming to a new set of coordinates $\left\{x^\alpha \right\}$ defined by,
$$x^0= y^0 \cosh y^1\;\;\; , \;\;\; x^1= y^0 \sinh y^1\;\;\; , \;\;\;x^2 = y^2  \;\;\; , \;\;\;x^3=y^3\eqno(37)$$
It is a simply connected space, which could be constructed with a topological identification of Minkowski space. In Misner coordinates, the required identification is
$$(y^0, y^1, y^2, y^3)\leftrightarrow (y^0, y^1+na, y^2, y^3)\eqno(38)$$
This is equivalent to the identification of timelike hypersurfaces $y^1 = na$, where $n$ is an integer. In Cartesian coordinates, it becomes
$$x^\alpha \equiv(x^0, x^1, x^2, x^3 )\leftrightarrow x_n^\alpha\equiv(x^0\cosh(na)+x^1\sinh(na),x^0\sinh(na)+x^1\cosh(na), x^2, x^3 )\eqno(39)$$
where it shows that the adjacent periodic boundaries are moving toward each other at a constant speed $v = \tanh a$  in the $x^1$ direction in the Minkowski space. To calculate $<0|T^{\mu\nu}|0>$, we need an appropriate vacuum state for Misner space. Although it does have local timelike Killing vector field everywhere, the nature of the topological identification is such that the local solutions cannot be patched together to form a global timelike Killing vector. However, each interval in Misner space is identical to a portion of Minkowski space and, provided the space-time is in the Minkowski vacuum state, a geodesic observer will not detect any particle. If we try to introduce local frames parallel to the coordinate axes, then the condition governing the identification of points forces us to Lorentz boost by velocity  $v=\tanh a$ to a new frame in the $x^1$  direction every time the Misner coordinate $y^1$ is increased by an amount $a$. This can be undone by the following condition on the spinor field
$$\Psi(x) = \left(\cosh({na\over2}) - \gamma^1\gamma^0\sinh({na\over2})\right)\Psi(x_n)\eqno(40)$$
for all integers $n$. The bracket behind $\Psi(x_n)$, induces the Lorentz boost in the $x^1$ direction at a speed $v = -\tanh(na)$. Expressed in terms of $G_{ren}^{(1)}$, it translates into
$$G_{ren}^{(1)}(\sigma) = \sum_{{n=-\infty}\atop{n \neq 0}}^\infty G_0^{(1)}(\sigma_n)\left[\cosh({na\over2}) + \gamma^1\gamma^0\sinh({na\over2})\right]\eqno(41)$$
The Misner space vacuum stress tensor calculation is straightforward. In the Misner coordinates the non-zero components are,
$$<0|T^{00}|0> = {M^4\over \pi^2}\sum_{n=1}^\infty \cosh({na\over2})\cosh(na)\left[{K_2(z_{n})\over z_{n}^2}\right]\eqno(42)$$
$$<0|T^{11}|0> = {M^4\over \pi^2}\sum_{n=1}^\infty \cosh({na\over2})\cosh(na){y^0}^2\left[{K_3(z_{n})\over z_{n}} - {K_2(z_{n})\over z_{n}^2}\right]\eqno(43)$$
$$<0|T^{22}|0> = <0|T^{33}|0> = -{M^4\over 2\pi^2}\sum_{n=1}^\infty \cosh({na\over2}){K_2(z_{n})\over z_{n}^2}\eqno(44)$$
where 
$$z_n = M|2y^0 \sinh({na\over 2})|\eqno(45)$$
Misner space contains nonchronal regions, where the roles of $y^0$ and $y^1$ are switched. The boundaries separating the chronal and nonchronal regions are null hypersurfaces, $x^0 = \pm x^1$, called {\it chronology horizons}. Examination of the vacuum stress-energy tensor shows that every component grows proportional to $(y^0)^{-4 }$ as $y^0$ approaches zero\footnote{To see this in more detail one could use the series expansion of the Bessel functions of the second kind [19] and then find the energy momentum components as $y^0 \rightarrow 0$.} at the chronology horizon. In the limit as $M\rightarrow 0$, the resulting components of the vacuum stress- energy tensor for a spinor field are
$$<0|T^{00}|0> = \sum_{n=1}^\infty  {1\over \pi^2}\cosh({na\over2})\cosh(na)\eqno(46)$$
$$<0|T^{11}|0> = \sum_{n=1}^\infty  {3{y^0}^2\over \pi^2}\cosh({na\over2})\cosh(na)\eqno(47)$$
$$<0|T^{22}|0> = <0|T^{33}|0> = -{1\over 2\pi^2} \sum_{n=1}^\infty \cosh({na\over2})\eqno(48)$$
The asymptotic form of the $<0|T^{\mu\nu}|0>$ components for the massive field, expanded about $y^0 = 0$, to the leading order are identical to the components given in equations (46-48). The ratio of each of these components for the massive field to that of the massless field approaches the unity near the chronology horizon. This ratio quickly decreases as $y^0$ increases, for the stress-energy components of the massive field decrease faster than those of the massless field. The ratio of the energy densities of massive to massless fields near $y^0 = 0$ is shown in Fig.5.\\
\begin{figure}
\begin{center}
\includegraphics[angle=-90,scale=0.7]{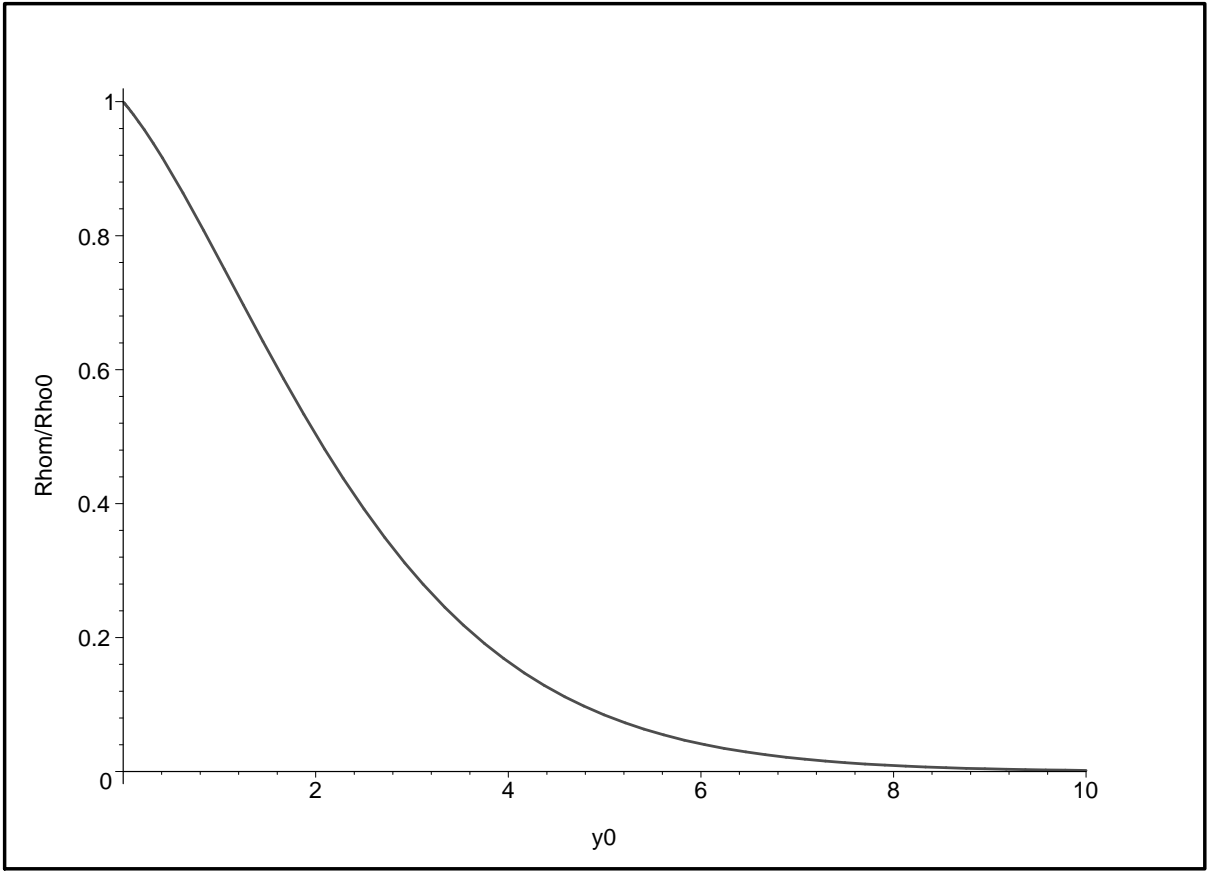}
\caption{The ratio of the spinor vacuum energy densities of a massive field to that of a massless field in Misner space. We have taken $M=a=1$.}
\end{center}
\end{figure}
\section{Conclusions}
Using the point separation regularization method we have calculated the vacuum stress-energy tensor for massive spinor fields in flat space-times with nontrivial topolgies including orientable and nonorientable manifolds. Both untwisted and twisted spin connections are taken into account It is shown that the spinor vacuum stress-energy tensor has the opposite sign and twice the magnitude of, the massive scalar stress-energy tensor. To consider the implications of spinor vacuum stress-energy tensor for the chronology protection conjecture we have calculated $<0|T^{\mu\nu}|0>$ for the Misner space-time which contains closed time like curves. It is shown that the vacuum stress-energy tensor diverges like $1\over {y^0}^4$ as $y^0\rightarrow 0$ at the chronology horizon. One should note that while it appears that the vacuum stress-energy tensor of quantized free fields does diverge as a chronology horizon is approached, the strength of the divergence may not be sufficient for the backreaction on the metric to prevent the formation of CTCs. However the ultimate mechanism for the chronology protection (assuming its existence), should the present attempts fail, is a propoerly constructed theory of quantum gravity.
\section*{Acknowledgement}
The authors would like to thank University of Tehran for supporting
this project under the grants provided by the research council.

\end{document}